\documentstyle[twocolumn,prl,aps,epsfig]{revtex}

\begin{document}
\draft
\input{psfig}
\title{Steady State of microemulsions in shear flow}
\author{F. Corberi}
\address{Istituto Nazionale per la Fisica della Materia,
Unit\`a di Salerno {\rm and} Dipartimento di Fisica, Universit\`a di Salerno,
84081 Baronissi (Salerno), Italy }
\author{G. Gonnella and D. Suppa}
\address{Istituto Nazionale per la  Fisica della Materia, Unit\`a di Bari
{\rm and} Dipartimento di Fisica, Universit\`a di Bari, {\rm and}\\
Istituto Nazionale di Fisica Nucleare, Sezione di Bari, via Amendola
173, 70126 Bari, Italy.}
\date{\today}
\maketitle
\begin{abstract}
Steady-state properties of microemulsions in shear flow
are studied in the context of a Ginzburg-Landau free-energy
approach. Explicit expressions are given  for 
the structure factor and the time correlation function at the
one loop level of approximation.
Our results predict
a four-peak pattern for the structure factor, implying the simultaneous
presence of interfaces aligned with two different orientations.
 Due to  the peculiar interface structure 
a non-monotonous  relaxation of the time correlator is also found.
\end{abstract}


Self-assembling amphiphilic systems are binary or ternary 
mixtures with a surfactant forming interfaces between the other 
fluid components.
These systems show a very 
rich phase behavior
with ordered and structured disordered
phases at different temperatures and  relative concentrations 
\cite{GR}.
Particularly  interesting  for the applications is 
the microemulsion phase where
coherent domains of oil and water 
on scales between 100 and 1000 $\AA$ form 
disordered isotropic intertwined structures on larger scales \cite{DeG}.

While a satisfactory comprehension of the equilibrium behavior of amphiphilic 
systems is now available \cite{GR}
 the dynamical features are far less understood.
In particular, in microemulsions, the existence of correlated mesoscopic 
structures makes the effects of imposed flows very different 
than in  simple fluids.
Interfaces between oil and water are 
elongated in the flow direction and an excess stress appears in the system.
This can give 
 a   peculiar dynamics with unusual relaxation properties
and a  non-Newtonian rheological behavior \cite{larson}.
Moreover, from a statistical
mechanics point of view, complex fluids in external flow are 
 an interesting example of an
 out of equilibrium
driven system \cite{zia}.

In this Letter we study the behavior of microemulsions in shear flow
 with a continuum free-energy approach similar 
to that used by Fredrickson and Larson \cite{Fre}
for the study of rheology
of block copolymers and by Onuki and Kawasaki \cite{onu} and others
\cite{miln} for the critical properties.
A Ginzburg-Landau model 
for describing
the rheological behavior of  ternary mixtures was 
considered
by P\"atzold and Dawson 
\cite{Daw1}. 
However the results of  \cite{Daw1} are based on an
evaluation of the structure factor 
obtained numerically or analytically only in the limit 
of vanishing shear rate $\gamma$.
Here we solve exactly the model for any value of $\gamma $ showing
that in an intermediate shear range
the structure factor is characterized by
four pronounced peaks in the plane
of the shear and flow directions. This 
new phenomenon can be interpreted in terms of 
a complex spatial pattern where interfaces with two different 
orientations coexist;
their relative abundance is tuned 
by $\gamma$. 
We  also consider the behavior 
of the two-time correlator.
Due to the coupling between the  flow and the interface structure,
this function has a  remarkable non monotonic behavior
characterized by the presence of 
maxima  preceding a fast
asymptotic decay.    
 
The  Ginzburg-Landau free-energy generally used
to describe the equilibrium properties of ternary mixtures is \cite{GS}
\begin{eqnarray}
{\cal F}\{\varphi\} = \int d^3 x
\{\frac{a}{2} \varphi^2 + \frac{b}{4!} \varphi^4 
&+& \frac{1}{2}(g_0 + g_2 \varphi^2) \mid \nabla \varphi \mid^2
\nonumber\\
&+&  \frac{c}{2} ( \nabla^2 )^2 \varphi \}
\label{eqn1}
\end{eqnarray}
where the scalar field $\varphi$ represents the concentration
difference between  oil and water components.
The amount of surfactant present in the system is related to 
the value of $g_0$ which can be negative, 
favoring the appearing of interfaces.
The term proportional to $c$ makes stable the free energy at large
momenta and weights the curvature of interfaces.
The quadratic part of the free-energy
in the disordered phase with  $a>0$ and $g_0^2 < 4 a c$ gives a suitable
description of the microemulsion phase; 
in particular the space correlation function 
\begin{equation}
G(r) \sim {\frac{e^{-r/\xi}}{r}} \sin{ \frac{2\pi r}{d}}
\label{eqn2}
\end{equation}
well fits the experimental data \cite{TS,W,GS2,others}. 
The characteristic length
$ d=2 \pi (1/2 \sqrt{a/ c}  - g_0 /4  c)^{-1/2}$
represents the size of coherent oil or water regions
which are correlated up to a distance 
$\xi= (1/2 \sqrt{a/ c}  +  g_0 /4 c)^{-1/2}$.
At negative $g_0$, 
the structure factor obtained by Fourier transforming Eq.~(2) 
has a peak at a finite value of the momentum given by $k_M=\sqrt{|g_0|/2c}$. 
Finally, the  non-linear terms in eq.(1)
describe the  possible effects due to mode coupling and become 
more important in approaching the phase boundary.

The kinetic behavior of the mixture, neglecting hydrodynamical
effects \cite{nota1}, is described 
by the convection-diffusion equation \cite{CH}
\begin{equation}   
\frac {\partial \varphi} {\partial t} + \vec \nabla (\varphi \vec v) =
\Gamma \nabla^2  \frac {\delta {\cal F}}{\delta \varphi} + \eta
\label{eqn2}
\end{equation}
where the velocity field is given by
\begin{equation}
\vec v = \gamma y \vec e_x
\label{eqn4}
\end{equation}
$\gamma$ being the  shear rate
and $\vec e_x$  the  unit vector in the flow direction.
The stochastic term 
$\eta$ is a gaussian white noise,  representing thermal fluctuations,
with zero mean and correlation
$\langle \eta(\vec r, t) \eta(\vec r', t')\rangle 
= -2 T \Gamma \nabla^2 \delta(\vec r -  \vec r') \delta(t-t')$,  
as required by the fluctuation-dissipation theorem,
where $\langle ...\rangle $ means the ensemble average. Here 
$\Gamma$ is a mobility coefficient and $T$  the 
temperature of the heat bath in contact with the fluid.

The cubic terms appearing  in 
the  functional derivative of  Eq. (3) will be treated in the one-loop  
approximation which  describes appropriately disordered phases. 
The resulting  equation for  the Fourier components 
$\varphi(\vec k, t)$ is given by 
\begin{eqnarray}
{\label{eqn5}}
\frac {\partial \varphi(\vec k,t)} {\partial t}&-& 
\gamma k_x \frac {\partial  \varphi(\vec k,t)} {\partial k_y}    
= -  \Gamma k^2 \left [ X(k)+g_2 k^2 S_0(t) \nonumber \right . \\
&+& \left . \left( \frac{b}{2}S_0(t)+ g_2 S_2(t) \right) \right ]
 \varphi(\vec k,t) +  \eta(\vec k, t) \hspace{.01cm}
\end{eqnarray}
where  $X(k)= c k^4 + g_0 k^2 +a$,
$S_0(t)$ and $S_2(t)$ are defined self-consistently
by the relations 
\begin{equation}
S_0(t)  =  \int _{|\vec k|<\Lambda}  \frac {d\vec k}{(2\pi)^D}  C(\vec k,t)
\end{equation}\begin{equation}
S_2(t)  =  \int _{|\vec k|<\Lambda}  \frac {d\vec k}{(2\pi)^D} k^2 C(\vec k,t)
\label{eqn6}
\end{equation} 
and  $\Lambda$ is  a phenomenological ultraviolet cut-off. 
Starting from  Eq.(5)
a straightforward calculation \cite {Daw1,corb} give 
the dynamical equations for the various correlation functions \cite{corr}.
We  first focus on the properties of the structure factor
$\langle \varphi(\vec k, t)\varphi(-\vec k, t)\rangle $.
 Using the method of characteristics
the $t \rightarrow \infty$ stationary expression 
 of the  structure factor  $C(\vec k)$ can be calculated as 
\begin{equation}
 C(\vec k) = \int_0^\infty dz 2 \Gamma T \vec K^2(z) 
e^{-\int_0^z ds 2\Gamma
{\vec K^2(s)}[a^R+g^R \vec K^2(s) +c \vec K^4(s)]}
\label{eqn6}
\end{equation}  
where $\vec{K}(s)= (k_x,k_y+ \gamma k_x s,k_z)$,  $ \vec K^4 
\equiv (\vec K^2)^2$,
$a^R=a+b/2S_0^\infty + g_2 S_2^{\infty}$, $g^R= g_0 + g_2 S_0^{\infty}$, and  
$S_0^{\infty}$ and $S_2^{\infty}$ are the limit for $t\to \infty $ of
$S_0(t)$ and $S_2(t)$.

\begin{figure}
\epsfig{file=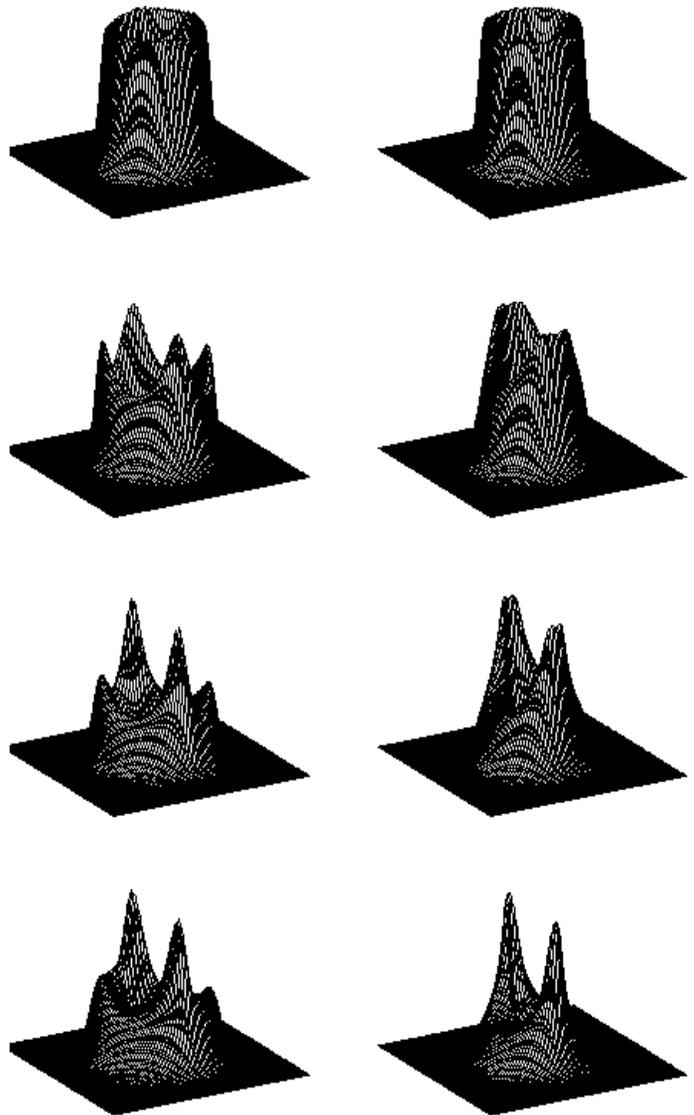,bbllx=70pt,bblly=45pt,bburx=553pt,bbury=800pt,
width=9.8 cm,height=16 cm,clip=}
\caption{The projections of the structure factor on the planes
$k_z = 0 $ (left column) and $k_y = 0 $ (right column) at shear rates
$\gamma = 0.25, 2, 4, 8$ (from top to bottom). 
The axis $k_x$ points towards the right. The vertical scale is the same
in all the pictures.} 
\end{figure}

Since $\xi$ and $d$ are the physical quantities to be compared with 
experimental data, the model parameters are tuned \cite{Daw1} 
with  the following procedure: We first calculate the one-loop 
renormalization of the bare coefficients $a$ and $g$ without shear 
and express the correlation lengths
$\xi$ and $d$ in terms of the renormalized parameters.
Then, for a given choice of $\xi$ and $d$,
we invert the above mentioned relations and compute
the values of $a,g_0$ to be used in the case with shear.
In the following we show results for the case $\xi=7, d=5$. 
The other parameters are set to the values
 $c=2$, $b = g_2 = T = 1$.
Similar results have been obtained for different choices of parameters
 with $d / \xi$ in the range between 0.5 and 2.5.
   
The effects of the flow on $C(\vec k)$ are  
shown in Fig.~1, where the 
projections of the three-dimensional structure factor on the planes
$k_y=0$ and $k_z=0$ are plotted at different $\gamma$. 
At $\gamma =0$ the structure factor is isotropic and its shape
on each cartesian
plane is that  of a circular volcano with radius $k_M$.
When shear is applied,  the projection on the $k_x=0$ plane
gives the same results of the $\gamma =0$ case because
the velocity field is  along the $x$ direction.
On the $k_z=0$ plane, instead, this pattern is
progressively distorted as the shear rate is increased.
For small $\gamma $ (see the case $\gamma =0.25$ in the left column of 
Fig.~1) the edge of the volcano becomes elliptical and  slightly 
depressed along the $k_x=-k_y$
direction so that two weak symmetric maxima 
located at $\theta= \arctan (k_y/k_x)=\pi/4, 3\pi/4$ are developed.
This is confirmed by a small $\gamma $ expansion of Eq.~(\ref{eqn6}).
As $\gamma $ is increased each of these two maxima 
is splitted into
two peaks, as shown in Fig.~1 at $\gamma=2$ and $\gamma =4$.
By  increasing the shear rate, the maxima located at 
$k_x\simeq 0$ become comparatively more important while the other peaks
rotate clockwise and decrease their amplitude linearly in $\gamma $
until they disappear.  Indeed, in
 the limit $\gamma \to \infty$, 
since terms proportional to powers of
$\gamma k_x$ damp the exponential term on the r.h.s. of Eq.~(\ref{eqn6}),
only the maxima of $C(\vec k)$ with $k_x=0$  survive.
 Fig.~1 shows that this is already observable
 for $\gamma =8$. Then, by letting
$k_x=0$ in Eq.~(\ref{eqn6})
 we find the peaks located at $k_y  = \pm k_M$.
The description  is completed with 
the projection of the structure factors on the plane
$ k_y = 0$ shown in the right column of Fig.~1.
Here two peaks at $k_x = 0, k_z = \pm k_M $, 
 are observed to become sharper and sharper as the
shear is increased. 

A peak of $C(k)$ is generally interpreted as the signature of
a characteristic length in the system which is proportional to the inverse
of its position. In this case, since the system is not isotropic,
to each maximum one associates three lengths, one for each space direction.
Due to the symmetry $\vec k\to -\vec k$ only the peaks  
 not related by reflection around the origin can be considered.
For large shear the existence of  a single couple of  maxima
at $k_x=0$ signals that interfaces are aligned along the flow.
In the transverse directions the characteristic lengths are the same as
without shear.
For intermediate values of $\gamma $ 
 the additional peaks at  $(\tilde k_x, \tilde k_y, \tilde k_z)$ 
reveal the 
presence of interfaces oriented with an angle 
$\alpha = \arctan (-\tilde k_x/ \tilde k_y)$ with respect to the flow, besides
those aligned along the $x$ direction. 
As $\gamma $ is
increased these features are progressively tilted in the direction 
of the flow while their relative abundance diminishes,
as suggested by the behavior of the maxima with $k_x\ne 0$
previously discussed. 
The existence of a four-fold peaked 
structure factor has been also reported in scattering experiments on
segregating mixtures \cite{quattropicchi,lcg}.  
For small $\gamma $ it is interesting to note that, although the 
microemulsion is almost isotropic, the depression of
$C(\vec k)$ along $k_x=-k_y$ indicates a slight predominance of
interfaces directed against the flow at $\alpha \simeq 3\pi/4$.

\begin{figure}
\epsfig{file=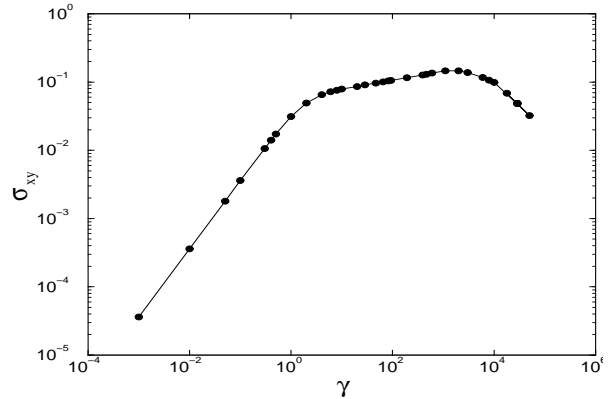,bbllx=30pt,bblly=40pt,bburx=530pt,bbury=500pt,
width=8 cm,height=6 cm,clip=}
\caption{The stress tensor as a function of the shear rate $\gamma$.}
\end{figure}

Stretching of domains requires work against surface tension and results
in an increase  $\Delta \eta$ of the viscosity \cite{Onu}. 
The excess viscosity is generally defined 
as  $\Delta \eta = \sigma_{xy} 
/ \gamma$  with the shear stress 
given by   
$\sigma_{xy} = -\int 
\frac {d\vec k}{(2\pi)^D} k_x k_y (g^R + 2 c k^2) C(\vec k) $
 \cite{Daw1}.
In fig.~2  $\sigma_{xy}$ is  plotted as a  
 function of $\gamma$. Shear thinning  is observed.
For small $\gamma $, $\sigma_{xy}$  grows linearly with  
$\gamma $   
as also  found in    \cite{Daw1}.
This behaviour is observed in correspondence of a structure factor
similar to that shown in Fig.~1 at $\gamma =0.25 $.
When  $C(\vec k)$ develops four maxima at  $k_z=0$,
$\sigma_{xy}$ keeps increasing with a much smaller
effective exponent  consistent with the value $0.13$.
For very large $\gamma$, when  the peaks at
$k_x=0$ alone survive, $\sigma_{xy}$ decreases to zero 
because the excess stress vanishes when the interfaces are 
aligned with the flow. A similar behavior is shown by
 the first normal
stress  $N_1 = \int 
\frac {d\vec k}{(2\pi)^D} (k_y^2- k_x^2)(g^R + 2 c k^2) C(\vec k)$
\cite{notan1}.

Next we consider the dynamical properties of the microemulsion phase.
In the steady state the two-time correlation function 
${\cal D}(\vec k,\vec k',t) = 
\langle \varphi(\vec k, t)\varphi(\vec k', 0)\rangle=
D(\vec k,t)\delta \left( \vec K(t)+k' \right )$ \cite{notadelta}
can be computed through 

\begin{equation}
{\label{eqn16}}
D(\vec k,t) =C\left(\vec K(t)\right)  
e^{- \int_0^t \Gamma
{\vec K^2(s)}[a^R+g^R \vec K^2(s) +c \vec K^4(s)] ds} 
\label{timecor}
\end{equation}

$D$ is the product of two terms: the structure factor
evaluated at the translated momentum $\vec K(t)$
and an exponential term.
For a given 
$\vec k = \vec K(0)$, $\vec K(t)$ moves along the $k_y$ direction upward
or downward depending on the sign of $k_x$, as represented 
schematically in the lower inset
of Fig.~3.
Therefore, when $\vec K(t)$ crosses the edge of the structure factor,
$C\left(\vec K(t)\right)$ shows a pronounced maximum. 
This feature is peculiar to microemulsions 
where the presence of interfaces gives the patterns 
of structure factor  discussed above
 and cannot be observed in simple fluids.
The exponential term in Eq.~(\ref{timecor}) contains a 
polynomial of $t$ and behaves as
$\exp (-\Gamma \gamma^6 k_x^6 c t^7/7)$ for long times.
However, due to the presence of negative coefficients,
the decay can be much slower at earlier times,
as it can be observed in the upper inset  of Fig.~3.
In particular one can show that an inflection point is developed 
at $\tan \theta = k_x/k_y=-\gamma t$ in the sectors with $k_x k_y<0$.
Other inflection points come out, even in the other sectors,
due to the negative value of $g^R$. Such a slow decay can preserve
the observation of the maximum of $C\left(\vec K(t)\right)$ in
the full time correlator (\ref{timecor}). 
In Fig.~3 the  behaviour of $D$ is shown for  three typical values
of $\vec k$. In  case A, $\vec K(t)$ does
not intersect the volcano of the structure factor so that $D$ decays
monotonously. When the edge of $C(\vec k)$ 
is crossed once or twice, as in cases B and C respectively, 
$D$ is characterized by a corresponding number of peaks.  
This rich relaxation behavior 
 has never been described before to our knowledge and 
could be searched for in experiments  \cite{notago}.
 
\begin{figure}
\epsfig{file=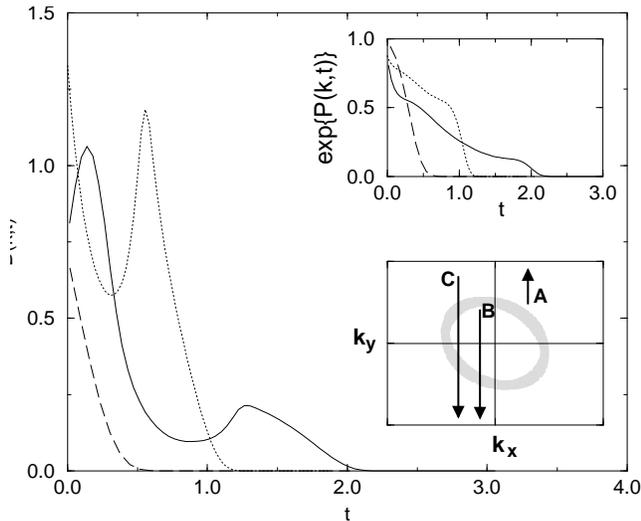,bbllx=60pt,bblly=40pt,bburx=570pt,bbury=440pt,
width=9.0 cm,height=7 cm,clip=}
\caption{The behavior of the time correlation function of Eq.~(9)
is shown for the three values of the wave-vector corresponding 
to the positions
A, B, C in the lower inset of the figure. The elliptical shape
in this inset represents schematically  the contour plot of the structure
factor in the plane $k_z=0$. Solid, dotted, and dashed lines correspond
respectively to the cases C, B, and A. 
In the upper inset the exponential factor
in the r.h.s. of Eq.~(9) is plotted for the same three values ok $\vec k$.}

\end{figure}

In conclusion, 
we have presented  explicit expressions for the steady-state
structure factor and the
 time correlation function of a Ginzburg-Landau model describing
mixtures of oil, water and surfactant in the microemulsion phase 
under the action of a shear flow.
The structure factor exhibits a four peak pattern which implies a non-uniform
orientation of interfaces with two preferred directions.
The time correlator, reminiscent of the interface structure,
shows maxima superimposed on a global
decay.
All these predictions can be tested  experimentally and show that
together with the equilibrium behavior also the dynamics of complex 
fluid is characterized by  a very rich phenomenology.
~\\
We thank Kenneth Dawson for helpful discussions. 
F.C. is grateful to M.Cirillo and R. Del Sole 
for hospitality in the University of Rome.
F.C. acknowledges support by the TMR network contract ERBFMRXCT980183
and by MURST(PRIN 97). 
G.G. acknowledges support by  MURST (PRIN  97).


\begin{references}
\bibitem{nota1}
We are also neglecting the motion of surfactant assuming 
that its relaxation is fast compared to that of the order parameter 
$\varphi$ (see \cite{Daw1}).
\end{references}
\end{document}